\documentclass[aps,pra,twocolumn,footinbib]{revtex4-1}
\usepackage{graphicx}
\usepackage{indentfirst}
\usepackage{braket}
\usepackage{float}
\usepackage{amsmath}
\usepackage{epstopdf}
\usepackage{CJK}
\usepackage{esint}
\usepackage{color}
\usepackage[T1]{fontenc}
\usepackage{subfigure}
\usepackage{amsfonts}
\usepackage{footmisc}
\usepackage{scrextend}
\usepackage{multirow}
\usepackage{cleveref}
\usepackage[english]{babel}
\usepackage{url}

\newcommand\argmax{\mathop{\mathrm{argmax}}}

\newcommand{\bp}{\boldsymbol \rho}

\def\be{\begin{equation}}
\def\ee{\end{equation}}
\def\ba{\begin{eqnarray}}
\def\ea{\end{eqnarray}}
\usepackage{bm}

\begin{document}

\title{Quantum illumination for enhanced detection of Rayleigh-fading targets}

\author{Quntao Zhuang$^{1,2}$}
\email{quntao@mit.edu}
\author{Zheshen Zhang$^1$}
\author{Jeffrey H. Shapiro$^1$}
\affiliation{$^1$Research Laboratory of Electronics, Massachusetts Institute of Technology, Cambridge, Massachusetts 02139, USA\\
$^2$Department of Physics, Massachusetts Institute of Technology, Cambridge, Massachusetts 02139, USA}
\date{\today}

\begin{abstract} 
Quantum illumination (QI) is an entanglement-enhanced sensing system whose performance advantage over a comparable classical system survives its usage in an entanglement-breaking scenario plagued by loss and noise.  In particular, QI's error-probability exponent for discriminating between equally-likely hypotheses of target absence or presence is 6\,dB higher than that of the optimum classical system using the same transmitted power.  This performance advantage, however, presumes that the target return, when present, has known amplitude and phase, a situation that seldom occurs in lidar applications.   At lidar wavelengths, most target surfaces are sufficiently rough that their returns are speckled, i.e., they have Rayleigh-distributed amplitudes and uniformly-distributed phases.  QI's optical parametric amplifier receiver---which affords a 3\,dB better-than-classical error-probability exponent for a return with known amplitude and phase---fails to offer any performance gain for Rayleigh-fading targets.   We show that the sum-frequency generation receiver [Phys. Rev. Lett. {\bf 118}, 040801 (2017)]---whose error-probability exponent for a nonfading target achieves QI's full 6\,dB advantage over optimum classical operation---outperforms the classical system for Rayleigh-fading targets.  In this case, QI's advantage is subexponential:  its error probability is lower than the classical system's by a factor of $1/\ln(M\bar{\kappa}N_S/N_B)$, when $M\bar{\kappa}N_S/N_B \gg 1$, with $M\gg 1$ being the QI transmitter's time-bandwidth product, $N_S \ll 1$ its brightness, $\bar{\kappa}$ the target return's average intensity, and $N_B$ the background light's brightness. 
\end{abstract} 

\maketitle

Quantum illumination (QI)~\cite{Sacchi_2005_1,Sacchi_2005_2,Lloyd2008,Tan2008,Lopaeva_2013,Guha2009,Ragy2014,Zheshen_15,Barzanjeh_2015} uses entanglement to 
outperform the optimum classical-illumination (CI) system for detecting the presence of a weakly-reflecting target that is embedded in a very noisy background, despite that environment's destroying the initial entanglement~\cite{footnote0}.  With optimum quantum reception, QI's error-probability exponent---set by the quantum Chernoff bound (QCB)~\cite{Audenaert2007}---is 6\,dB higher~\cite{Tan2008} than that of the optimum CI system, i.e., a coherent-state transmitter and a homodyne receiver.  Until recently, the sole structured receiver for QI that outperformed CI---Guha and Erkmen's optical parametric amplifier (OPA) receiver~\cite{Guha2009}---offered only a 3\,dB increase in error-probability exponent.  In Ref.~\cite{Zhuang_2017}, we showed that the sum-frequency generation (SFG) receiver's error-probability exponent reached QI's QCB.  Moreover, augmenting that receiver with feed-forward (FF) operations yielded the FF-SFG receiver~\cite{Zhuang_2017}, whose performance, for a low-brightness transmitter, matched QI's Helstrom limit for both the target-detection error probability and the Neyman-Pearson criterion's receiver operating characteristic (ROC)~\cite{zhuang2017entanglement}.  

Prior QI performance analyses~\cite{Tan2008,Guha2009,Zhuang_2017,zhuang2017entanglement} have all assumed that the target return has known amplitude and phase, something that seldom occurs in lidar applications.  At lidar wavelengths, most target surfaces are sufficiently rough that their returns are speckled, i.e., they have Rayleigh-distributed amplitudes and uniformly-distributed phases~\cite{Goodman1965,Goodman1976,Shapiro1981,Shapiro1982}.  It is crucial, therefore, to show that QI maintains a target-detection performance advantage over CI for a target return with random amplitude and phase.  

In this paper, we compare QI and CI target detection for Rayleigh-fading targets in the flat-fading limit, when the complex-field envelope of the target return from a single transmitted pulse suffers multiplication by a time-independent Rayleigh-distributed random amplitude and a time-independent uniformly-distributed random phase shift.  We show that QI with OPA reception fails to offer any performance advantage over CI in this case.  QI with SFG reception does provide an advantage over CI:  when $M\bar{\kappa}N_S/N_B \gg 1$, its error probability is a factor of $1/\ln(M\bar{\kappa}N_S/N_B)$ lower than that of optimum CI, which transmits a coherent state and uses heterodyne reception.  Here, $M\gg 1$ is the QI transmitter's time-bandwidth product, $N_S$ is its brightness, $\bar{\kappa}$ is the target return's average intensity, and $N_B$ is the background light's brightness.     

{\em QI target detection}---.
In QI, the transmitter illuminates the region of interest with a single-spatial-mode, $T$-s-long pulse of signal light produced by pulse carving the continuous-wave output of a spontaneous parametric downconverter (SPDC).  The SPDC source is taken to have a $W$-Hz-bandwidth, flat-spectrum phase-matching function with $W \gg 1/T$.  The resulting signal pulse is maximally entangled with a corresponding single-spatial-mode, $T$-s-long pulse of idler light that the transmitter retains for subsequent joint measurement with the light returned from the region of interest.  The $M = TW \gg 1$ signal-idler mode pairs that comprise the transmitted signal and retained idler pulses are thus in independent, identically-distributed (iid), two-mode squeezed-vacuum states with average photon number $N_S \ll 1$ in each signal and idler mode.  Let $\{\hat{a}_{S_m},\hat{a}_{I_m}\}$ be the photon-annihilation operators for the transmitter's $M$ signal and idler modes, and $\{\hat{a}_{R_m}\}$ the photon-annihilation operators of the $M$ modes returned from the region of interest. The target-detection hypothesis test is to determine whether $h=0$ (target absent) or $h=1$ (target present) is true when:  $\hat{a}_{R_m} = \hat{a}_{B_m}$, for $h =0$, and $\hat{a}_{R_m} = \sqrt{\kappa}\,e^{i\phi}\hat{a}_{S_m} + \sqrt{1-\kappa}\,\hat{a}_{B_m}$, for $h=1$.  Here: the $\{\hat{a}_{B_m}\}$ are photon-annihilation operators for iid background-noise modes that are in the thermal state with average photon number $N_B \gg 1$ when $h=0$ and in the thermal state with average photon number $N_B/(1-\kappa)$ when $h=1$~\cite{footnote1}; $\kappa > 0$ is the target-return's reflectivity; and $\phi$ is the target-return's phase.  

Previous theoretical work on QI target detection~\cite{Tan2008,Guha2009,Barzanjeh_2015,Zhuang_2017} has assumed known $\kappa$, $\phi = 0$~\cite{footnote2},
and lossless idler storage.  For equally-likely target absence or presence, QI with optimum quantum reception---realizable with FF-SFG~\cite{Zhuang_2017}---has error probability $\Pr(e)_{\rm opt} \simeq e^{-M\kappa N_S/N_B}/2$, QI with OPA reception has error probability $\Pr(e)_{\rm OPA} \simeq e^{-M\kappa N_S/2N_B}/2$, and optimum CI has error probability $\Pr(e)_{\rm CI} \simeq e^{-M\kappa N_S/4N_B}/2$.  

Lidar targets are almost always speckle targets, viz., $\sqrt{\kappa}$ and $\phi$ are statistically independent random variables whose respective probability density functions (pdfs) are
$f_{\sqrt{\kappa}}(x) = 2xe^{-x^2/\bar{\kappa}}/\bar{\kappa}$, for $x >0$, and 
$f_\phi(y) = 1/2\pi$, for $0\le y\le 2\pi$,
where $\bar{\kappa}$ is the target return's average intensity.  These statistics invalidate \emph{all} of the error-probability expressions from the preceding paragraph.  Worse, as will soon be seen, they preclude \emph{any} QI receiver from obtaining a single-pulse error probability that decreases exponentially with increasing $M\bar{\kappa}N_S/N_B$.  For that demonstration we will employ the QCB, an exponentially-tight upper bound on the error probability of optimum quantum reception for multiple-copy quantum state discrimination~\cite{Audenaert2007}.  

{\em The QCB applied to QI with Rayleigh fading}---.
Conditioned on knowledge of $h$, $\sqrt{\kappa}$, and $\phi$, the $\{\hat{a}_{R_m},\hat{a}_{I_m}\}$ mode pairs at the QI receiver are in the state $\hat{\bp}_h(\sqrt{\kappa},\phi) = \otimes_{m=1}^M\hat{\rho}^{(m)}_h(\sqrt{\kappa},\phi)$, with $\hat{\rho}_h^{(m)}(\sqrt{\kappa},\phi)$ being the two-mode, zero-mean, Gaussian state whose  Wigner covariance matrix is 
\ba
&
{\mathbf{\Lambda}}_h =
\frac{1}{4}
\left[
\begin{array}{cccc}
(2N_B+1) {\mathbf I}&2C_p{\mathbf R}_h\\
2C_p{\mathbf R}_h&(2N_S+1){\mathbf I}
\end{array} 
\right],  
\label{hk}
&
\ea
where $N_B \gg 1\gg N_S$ has been used.  In this covariance matrix:   ${\bf I}$ is the $2\times 2$ identity matrix, and ${\mathbf R}_h={\rm Re}\!\left[e^{i\phi} \left({\mathbf Z}-i{\mathbf X}\right)\right]\delta_{h1}$, where $\delta_{hk}$ is the Kronecker delta function, and ${\mathbf Z}$ and ${\mathbf X}$ are $2\times 2$ Pauli matrices.   It follows that the signature of target presence is the nonzero phase-sensitive cross correlation, $C_p=\sqrt{\kappa N_S\left(N_S+1\right)}$, between the returned signal and the retained idler modes.

Erroneous target-detection decisions can be either false-alarm errors, when target presence is declared but no target is present, or miss errors, when target absence is declared but a target is present.  For a given target-detection system, the conditional probabilities for these errors to occur are the false-alarm probability $P_F$, and the miss probability $P_M = 1-P_D$, where $P_D$ is the detection probability, i.e., the probability that target presence is declared when a target is present.  Almost all QI target detection analyses~\cite{Tan2008,Guha2009,Barzanjeh_2015,Zhuang_2017} have been Bayesian:  assign prior probabilities, $\{\pi_h\}$, to $h=0$ and $h=1$, and minimize the error probability, $\Pr(e) = \pi_0P_F + \pi_1P_M$, typically for equiprobable hypotheses, $\pi_0 = \pi_1 = 1/2$.  Owing to the difficulty of accurately assigning priors to target absence and presence, a better approach to optimizing target-detection performance is to apply the Neyman-Pearson performance criterion: maximize $P_D$ subject to a constraint on $P_F$.  Only recently has this criterion been applied to QI target detection~\cite{zhuang2017entanglement}, and that work assumed knowledge of the target return's amplitude and phase.  In this paper, we will consider both performance criteria---minimizing $\Pr(e)$ and maximizing $P_D$ for a given $P_F$---for our Rayleigh-fading QI scenario.

In the Bayesian approach, the minimum error probability for QI target detection is set by the Helstrom limit~\cite{Helstrom1969} for discriminating between the \emph{unconditional} $h=0$ and $h=1$ states, 
\begin{equation}
\hat{\bar{\bp}}_h = \int\!{\rm d}x\int\!{\rm d}y\,f_{\sqrt{\kappa}}(x)f_\phi(y)\hat{\bp}_h(x,y).  
\end{equation}
This limit's calculation requires diagonalizing $\pi_1\hat{\bar{\bp}}_1-\pi_0\hat{\bar{\bp}}_0$, so it is intractable for QI with Rayleigh fading,  because $\hat{\bar{\bp}}_1$ is not an $M$-fold product state.  Nevertheless, applying the QCB will yield an informative result.

Let $D_{\pi_0}(\hat{\bp}_0(x,y),\hat{\bp}_1(x,y))$ denote the Helstrom limit for discriminating between $\hat{\bp}_0(x,y)$ and $\hat{\bp}_1(x,y)$ that occur with priors $\pi_0$ and  $\pi_1$, and let $\xi_{\rm QCB}(\hat{\bp}_0(x,y),\hat{\bp}_1(x,y)) \equiv -\lim_{M\rightarrow \infty} {\ln[D_{\pi_0}(\hat{\bp}_0(x,y),\hat{\bp}_1(x,y))]}/M$ be the QCB on its error-probability exponent.   Then, using the Helstrom limit's being concave in quantum states (see Lemma~1 in the Appendix), we can show (see Lemma~2 in the Appendix) that the Helstrom limit's error-probability exponent for QI target detection, $\xi_{\rm QI}\equiv-\lim_{M\to\infty}{\ln[D_{\pi_0}(\hat{\bar{\bp}}_0,\hat{\bar{\bp}}_1)]}/M$, vanishes, i.e., $\xi_{\rm QI} = 0$, for all $\pi_0\pi_1 \neq 0$.  Having $\xi_{\rm QI} = 0$ implies that optimum quantum reception for QI target detection with Rayleigh fading has an error probability that decreases subexponentially with the number of signal-idler mode pairs that are employed.  This subexponential error-probability behavior applies to \emph{all} QI receivers, including the FF-SFG, SFG, and OPA receivers.  Because OPA receivers are relatively easy to build~\cite{Zheshen_15}---as opposed to the far more complicated SFG and FF-SFG receivers~\cite{Zhuang_2017}---one might hope that QI with OPA reception would offer a performance advantage over optimum CI for the Rayleigh-fading scenario.  We next show that such is not the case.  

{\em OPA reception for QI with Rayleigh fading}---.  
It is difficult to get an analytic error-probability approximation for QI with OPA reception in the Rayleigh-fading scenario, so we will content ourselves with finding its SNR and comparing that result to the SNR for the optimum Rayleigh-fading CI system.  
The OPA receiver's essence is converting QI's phase-sensitive cross-correlation signature of target presence to an average photon-number signature that can be sensed with direct detection.  
In particular, the OPA receiver measures
$\hat{N} \equiv\sum_{m=1}^M \hat{a}_m^\dagger\hat{a}_m$, where
$\hat{a}_m = \sqrt{G}\,\hat{a}_{I_m} + \sqrt{G-1}\,\hat{a}_{R_m}^\dagger$
is the idler-port output of a low-gain ($\max(N_S/N_B,N_S/\kappa N_B^2) \ll G-1 \sim \sqrt{N_S}/N_B \ll 1$) OPA.  Hence, we define its SNR to be
${\rm SNR}_{\rm OPA} \equiv [(\sum_{j=0}^1(-1)^j\langle\hat{N}\rangle_j) /(\sum_{j=0}^1\sqrt{{\rm Var}_j(\hat{N})})]^2$,
where $\langle \hat{N}\rangle_j$ and ${\rm Var}_j(\hat{N})$ for $j=0,1$ are the conditional means and conditional variances of the $\hat{N}$ measurement given $h=j$.

For known $\kappa$ and $\phi =0$, we get
$\langle\hat{N}\rangle_1 - \langle\hat{N}\rangle_0 \approx 2M\sqrt{G(G-1)\kappa N_S(N_S+1)}$.  Combining this result with ${\rm Var}_j(\hat{N}) \approx \langle\hat{N}\rangle_j$ for the $\hat{N}$ measurement's conditional variances, gives ${\rm SNR}_{\rm OPA} \approx M\kappa N_S/N_B$ when $N_S \ll 1$, $\kappa\ll 1$ is known, $\phi=0$, and $N_B \gg 1$.  In the Rayleigh-fading case, the uniformly-distributed random phase destroys the phase-sensitive cross-correlation signature in $\langle \hat{N}\rangle_1$, leading to $\langle\hat{N}\rangle_1 - \langle\hat{N}\rangle_0 = M(G-1)\bar{\kappa}N_S$, and it adds $2M^2(G-1)\bar{\kappa}N_S$ to ${\rm Var}_1(\hat{N})$, hence giving us
\begin{equation}
{\rm SNR}_{\rm OPA} \approx \frac{M(G-1)(\bar{\kappa}N_S)^2/N_B}{(1+\sqrt{1+ 2M\bar{\kappa}N_S/N_B})^2},
\end{equation}
which is much smaller than $M\bar{\kappa}N_S/N_B$, the ${\rm SNR}_{\rm OPA}$ for a known $\kappa = \bar{\kappa}$ and $\phi=0$~\cite{footnote3}.  

Optimum CI for Rayleigh fading does matched filtering of its heterodyne detector's output followed by square-law envelope detection that yields an output, $R$, which is exponentially distributed under both $h=0$ and $h=1$~\cite{VanTrees1}.  The SNR for this system, ${\rm SNR}_{\rm CI} \equiv [(\sum_{j=0}^1(-1)^j\langle R\rangle_j)/(\sum_{j=0}^1\sqrt{{\rm Var}_j(R)})]^2$, satisfies
\begin{equation}
{\rm SN}_{\rm CI} = (M\bar{\kappa} N_S/2N_B)/\left(1+M\bar{\kappa}N_S/2N_B\right)^2,
\end{equation}
which is orders of magnitude greater than ${\rm SNR}_{\rm OPA}$ for Rayleigh fading in the interesting $M\bar{\kappa}N_S/N_B \gg 1$ operating regime.

\begin{figure}[th]
\includegraphics[width=0.4\textwidth]{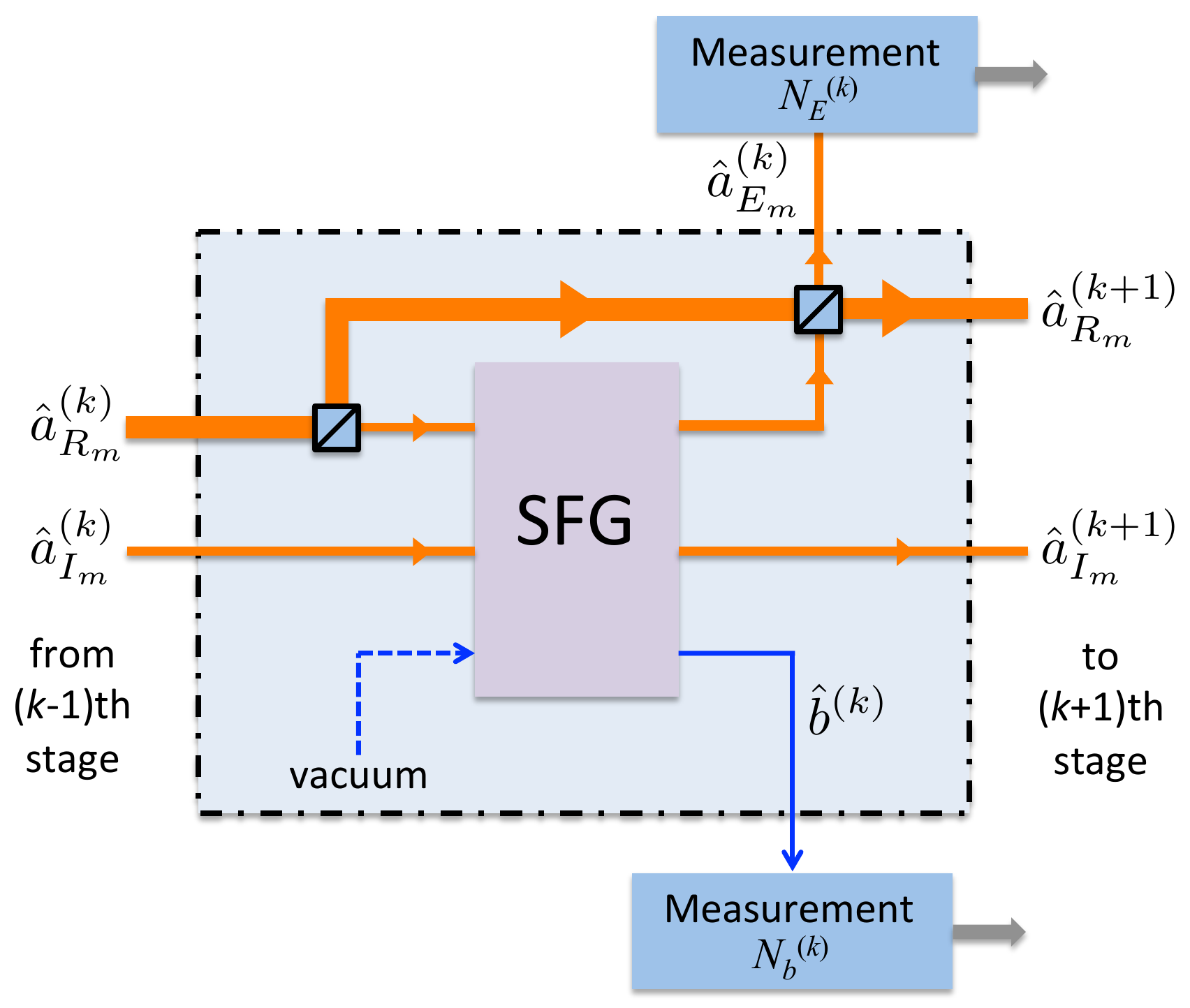}
\caption{Schematic representation of the sum-frequency generation (SFG) receiver's $k$th stage, showing only the $m$th mode pair, although all $M$ mode pairs are processed simultaneously.  The $m$th mode pair of the returned light ($\hat{a}_{R_m}^{(k)}$) and the retained idler ($\hat{a}_{I_m}^{(k)}$) at the input to the $k$th stage is transformed into the corresponding mode pair at that stage's output by means of SFG.  Photon-counting measurements are made on the single-mode sum-frequency output ($\hat{b}^{(k)}$) and the auxiliary output modes ($\{\hat{a}_{E_m}^{(k)} : 1 \le m \le M\}$).  The SFG receiver's decision as to target absence or presence is based on the total of all the photon-counting measurements, i.e., $N_T \equiv \sum_{k=1}^K(N_b^{(k)} + N_E^{(k)}),$ where $N_b^{(k)}$ is the outcome of the $\hat{b}^{(k)\dagger}\hat{b}^{(k)}$ measurement, and $N_E^{(k)}$ is the outcome of the $\sum_{m=1}^M\hat{a}_{E_m}^{(k)\dagger}\hat{a}_{E_m}^{(k)}$ measurement.} 
\label{SFGrcvr}
\end{figure}

{\em SFG Reception for QI with Rayleigh Fading}---.
The SFG receiver~\cite{Zhuang_2017} uses a succession of $K$ SFG stages.  At the input to each such stage a beam splitter taps off a small fraction of the light returned from the region of interest to undergo SFG with the retained idler light.  The returned-light output from that SFG process is then recombined with the portion remaining from that stage's input beam-splitter and applied, along with the retained-idler output, to the next stage.  Photon-counting measurements are performed on the SFG's sum-frequency output and on the auxiliary output from the return-light beam splitter at the output of each SFG stage.  These measurements are used to decide on target absence or presence.  Figure~\ref{SFGrcvr} shows a schematic representation of the SFG receiver's $k$th stage, for more details see Ref.~\cite{Zhuang_2017}.  

For known $\kappa$ and $\phi = 0$, SFG reception's error probability achieves the QCB.  The FF-SFG receiver~\cite{Zhuang_2017} augments the SFG receiver with pre-SFG and post-SFG squeezers, whose parameters are chosen in accordance with a Bayesian update rule that is controlled by feed-forward information from the prior stages.  FF-SFG reception reaches the Helstrom limit for QI target detection---in both the Bayesian and Neyman-Pearson settings---for known $\kappa$ and $\phi = 0$~\cite{Zhuang_2017,zhuang2017entanglement}.  Because its feed-forward operations exploit $\phi=0$, FF-SFG reception ceases to function effectively when $\phi$ is uniformly distributed.  SFG reception, which eschews the use of feed-forward, \emph{does} cope with random amplitude and phase, as we now show.   

When $h=0$, the SFG receiver's total photon count---i.e., $N_T\equiv \sum_{k=1}^K(N_b^{(k)} + N_E^{(k)})$ from Fig.~\ref{SFGrcvr}---is the sum of $M$ iid Bose-Einstein random variables, and has mean value $N_0 \simeq -N_S \ln(\epsilon)/2$ for $N_S \ll 1$.   When $h=1$, and conditioned on the values of $\kappa$ and $\phi$, the statistics of the SFG receiver's total photon count equal those for direct detection of the coherent state $|\sqrt{(1-\epsilon)M\kappa N_S/N_B}\,e^{i\phi}\rangle$ embedded in a weak thermal-noise background of average photon number $N_0 \ll 1$.  In these expressions, $\epsilon \ll 1$ is chosen to obtain good performance, see~\cite{Zhuang_2017} for details.  When $M\kappa N_S/N_B \gg N_0$, the thermal contribution to the $h=1$ statistics can be neglected.  Then, averaging the $h=1$ conditional state over the $\sqrt{\kappa}$ and $\phi$ statistics results in a thermal state with average photon number $N_1 = (1-\epsilon)M\bar{\kappa}N_S/N_B$, implying that the SFG receiver has reduced Rayleigh-fading QI target detection to discriminating between two thermal states, 
$\hat{\sigma}_{0} = \sum_{n=0}^\infty [N_0^n/(N_0+1)^{(n+1)}]\ket{n}\bra{n}$ and  
$\hat{\sigma}_{1} = \sum_{n=0}^\infty [N_1^n/(N_1+1)^{(n+1)}]\ket{n}\bra{n}$, using photon-counting measurements.
SFG reception's minimum error-probability decision, $\tilde{h} = 0$ or 1, is therefore  
$\tilde{h}=\argmax_h \pi_h \!\left[N_h^n/(N_h+1)^{(n+1)}\right]$, where $n$ is the observed photon count.  

The preceding rule can be implemented as a threshold test:  $\tilde{h} = 1$ if and only if $n > n_t$, where the threshold $n_t$ satisfies $\pi_0 N_0^{n_t}/(N_0+1)^{(n_t+1)}\ge \pi_1 N_1^{n_t}/(N_1+1)^{(n_t+1)}$ and $\pi_0 N_0^{n_t+1}/(N_0+1)^{(n_t+2)}< \pi_1 N_1^{n_t+1}/(N_1+1)^{(n_t+2)}$.  SFG reception's ROC---its $P_D$ versus $P_F$ behavior---can now be obtained analytically.  For integer $n_t$, we have $P_F^{\rm SFG}= [N_0/(N_0+1)]^{n_t+1}$ and $P_D^{\rm SFG}=[N_1/(N_1+1)]^{n_t+1}$.  ROC points intermediate between those generated with integer thresholds are then obtained from randomized tests~\cite{VanTrees2}.

The Bayesian approach's error probability is easily found once its decision rule's threshold $n_t$ is determined.  Evaluating the false-alarm and detection probabilities for that threshold value, SFG reception's error probability then follows from 
$
\Pr(e)_{\rm SFG}=\pi_0P_F^{\rm SFG}+\pi_1 (1-P_D^{\rm SFG}).
$
For $N_S\to 0$ with $\epsilon\ll1$, we find that $n_t=0$ and hence
\be
\Pr(e)_{\rm SFG}\simeq \Pr(e)_{\rm SFG}^{N_S\to0}\equiv  \pi_1/(1+M\bar{\kappa}N_S/N_B).
\label{NS0}
\ee
This result's algebraic scaling with $M$ is consistent with our earlier finding that optimum quantum reception for Rayleigh-fading QI target detection has an error probability that decreases subexponentially with increasing $M$.  

\begin{figure}
\includegraphics[width=0.225\textwidth]{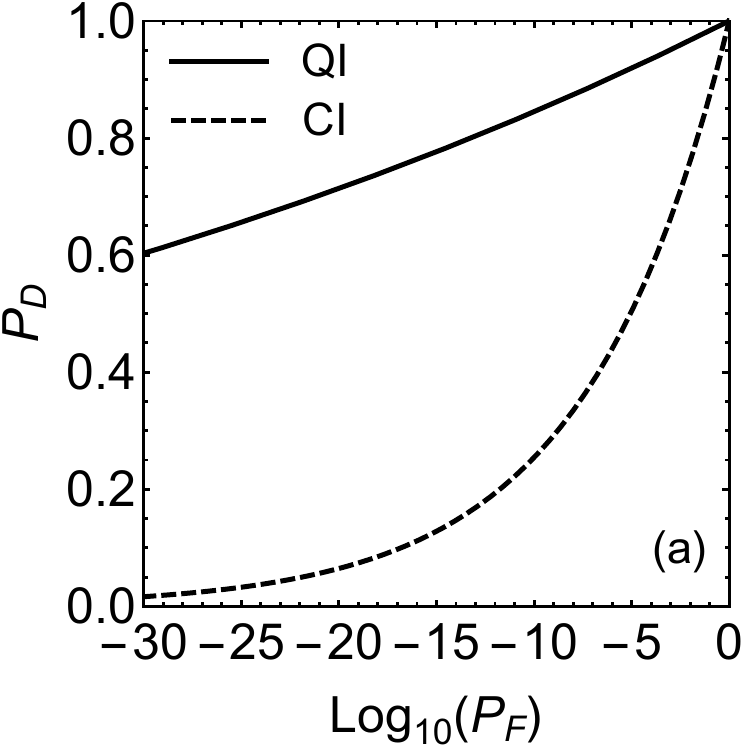}\hspace*{.1in}
\includegraphics[width=0.225\textwidth]{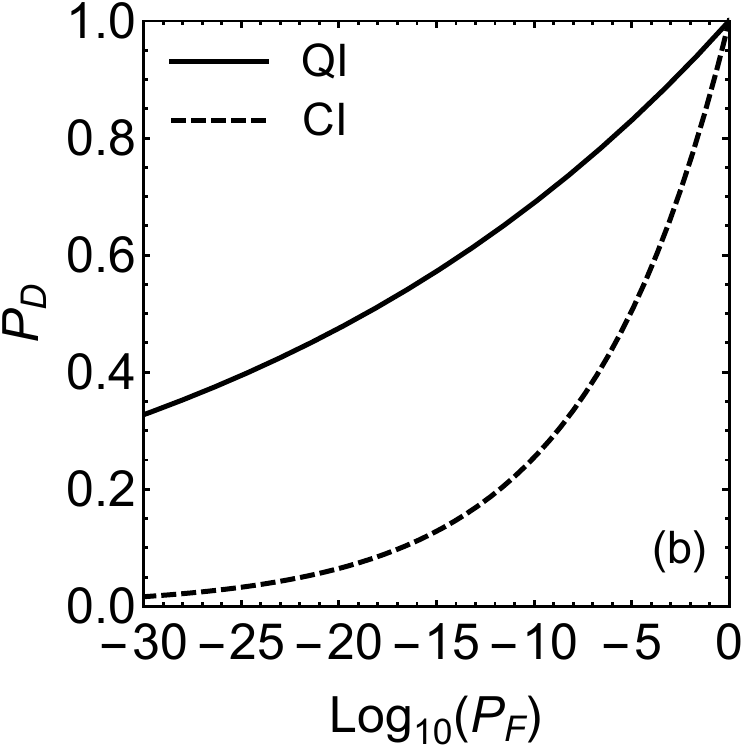}
\caption{QI and CI ROCs for Rayleigh-fading target detection with $\bar{\kappa}=0.01$, $N_B = 20$, and $\epsilon=0.01$. (a) $N_S=10^{-4}$ and $M=10^{8.5}$. (b) $N_S=10^{-2}$ and $M=10^{6.5}$.
\label{Fig_comparisons_ROC}
}
\end{figure}

{\em QI versus CI for Rayleigh Fading}---.
We are now prepared to demonstrate that QI target detection with SFG reception enjoys a significant performance advantage over CI target detection in the Rayleigh-fading scenario.  We start with the Neyman-Pearson criterion, for which we already have the ROC for QI with SFG reception.  The ROC for CI target detection with a coherent-state transmitter and heterodyne detection is~\cite{VanTrees1} 
$P_D^{\rm CI}=\left(P_F^{\rm CI}\right)^{1/{\left(1+M\bar{\kappa}N_S/N_B\right)}}.$
Figure~\ref{Fig_comparisons_ROC} compares two QI and CI ROCs.  Similar to what was assumed in Refs.~\cite{Tan2008,Zhuang_2017}, we took $\bar{\kappa} = 0.01$, $N_B=20$, and $\epsilon=0.01$ for both comparisons.  In one case we assumed $N_S = 10^{-4}$ and $M = 10^{8.5}$, while in the other we chose $N_S = 10^{-2}$ and $M = 10^{6.5}$.  Figure~\ref{Fig_comparisons_ROC} shows that QI target detection with SFG reception has a much higher detection probability than optimum CI target detection at low false-alarm probabilities.

\begin{figure}
\includegraphics[width=0.225\textwidth]{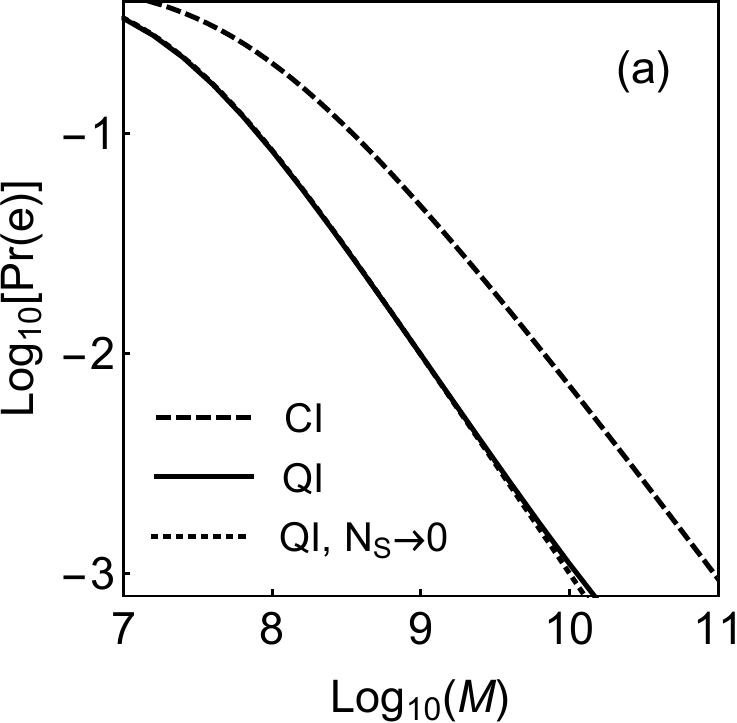}
\hspace*{.1in}
\includegraphics[width=0.225\textwidth]{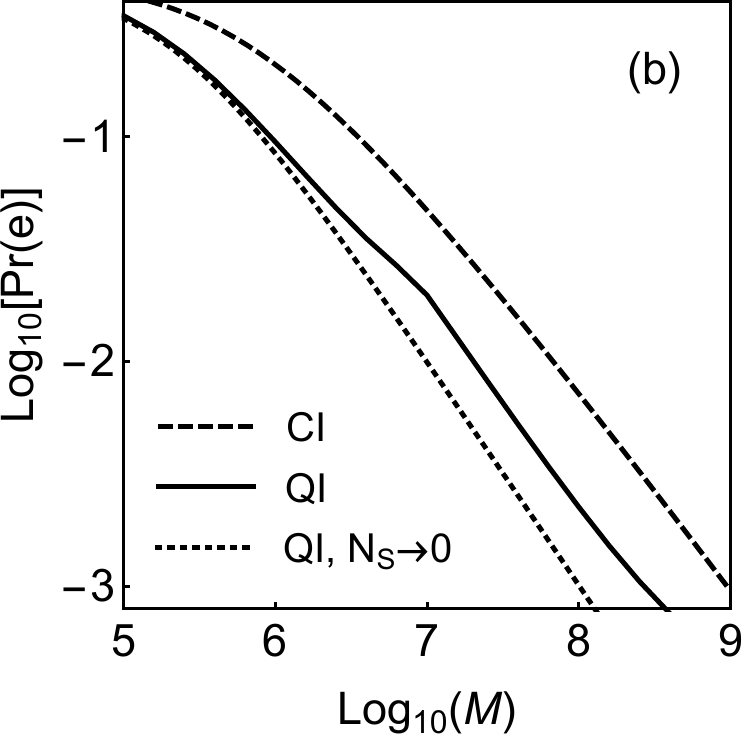}
\caption{QI and CI error probabilities for Rayleigh-fading target detection with $\pi_0=\pi_1=1/2$, $\bar{\kappa} = 0.01$, $N_B=20$, and $\epsilon=0.01$. (a) $N_S=10^{-4}$. (b) $N_S=10^{-2}$. The slope discontinuity in $\Pr(e)_{\rm SFG}$ for $N_S = 10^{-2}$ is due to the its receiver's photon-number threshold increasing from $n_t = 0$ to $n_t = 1$ at that point.
\label{Fig_comparisons}
}
\end{figure}

Turning now to the Bayesian approach, we again have the QI result in hand, and we find optimum CI's error probability from $\Pr(e)_{\rm CI}=\min_{P_F^{\rm CI}}[\pi_0 P_F^{\rm CI}+\pi_1(1-P_D^{\rm CI})]$. Figure~\ref{Fig_comparisons} plots $\Pr(e)_{\rm SFG}$ and $\Pr(e)_{\rm CI}$ versus $\log_{10}(M)$ for equally-likely target absence or presence assuming $\bar{\kappa} = 0.01$, $N_B = 20$, and $\epsilon = 0.01$ for $N_S = 10^{-4}$ and $N_S = 10^{-2}$.  Here we see that QI target detection with SFG reception offers a significantly lower error probability than optimum CI target detection.  Indeed, for $MN_S \gg 1$ we obtain the asymptotic result
\be
\Pr(e)_{\rm CI}\simeq  \frac{\pi_1\ln(M\bar{\kappa}N_S/N_B)}{M\bar{\kappa}N_S/N_B}+O\!\left(\frac{1}{MN_S}\right),
\ee
which is a factor of $\ln(M\bar{\kappa}N_S/N_B)$ higher than the corresponding result for $\Pr(e)_{\rm SFG}^{N_S\to0}$ when $M\bar{\kappa}N_S/N_B \gg 1$.  Moreover, Fig.~\ref{Fig_comparisons}a shows that $N_S = 10^{-4}$ is small enough to ensure $\Pr(e)_{\rm SFG} \approx \Pr(e)_{\rm SFG}^{N_S\to0}$ for the parameter values employed therein.  At high enough $M$ values, however, the effect of background noise in the SFG process becomes significant and $\Pr(e)_{\rm SFG}$ begins to deviate from the ideal $N_S\to0$ result.  The onset of this deviation occurs at lower $M$ values when $N_S=10^{-2}$, as seen in Fig.~\ref{Fig_comparisons}b, because the background-noise effect on the SFG process is proportional to $N_S$~\cite{Zhuang_2017}. Nevertheless, QI's advantage over CI persists. We also see that QI target detection's robustness to noise is worse for Rayleigh fading than what our previous results~\cite{Zhuang_2017} showed for known $\kappa$. This reduced robustness arises from noise having  greater impact on Rayleigh-fading error probability---because $\kappa \ll \bar{\kappa}$ can occur---as opposed to its effect in a nonfading environment with $\kappa = \bar{\kappa}$.   

{\em Conclusions}---.
QI target detection is remarkable because it uses entanglement to outperform CI despite environmental loss and noise's destroying that entanglement.  Previously, both theory and experiment have demonstrated QI's having an advantage over CI, but \emph{only} for a target return with known amplitude and known phase.  Yet lidar targets are generally speckle targets, so  their target returns have Rayleigh-distributed amplitudes and uniformly-distributed phases.  We have shown that SFG reception affords a target-detection performance advantage over optimum CI for this scenario, but its magnitude is much smaller than what QI provides for the nonfading situation.  Nevertheless, our result brings QI target detection closer to practical application, although two major problems remain to be solved:  implementing near-lossless idler-storage and near-unity efficiency SFG for low-brightness, broadband light.  

Two final points now deserve mention.  First, although we have limited our treatment to the Rayleigh-fading scenario, the SFG receiver's immunity to a uniformly-distributed random phase means that it will also be effective against other fading distributions, e.g., the Rician fading that models a target return with both specular and diffuse components~\cite{VanTrees1,swerling1997radar}.  Finally, because $N_B\gg 1$ most naturally occurs at microwave, rather than optical, wavelengths~\cite{Barzanjeh_2015}, SFG reception's applicability to a variety of flat-fading scenarios makes it relevant for microwave as well as optical QI.  

Q.~Z. acknowledges support from the Claude E. Shannon Research Assistantship. Z.~Z. and J.~H.~S. acknowledge support from Air Force Office of Scientific Research Grant No.~FA9550-14-1-0052.

\setcounter{equation}{0}
\makeatletter
\renewcommand{\theequation}{A\arabic{equation}}
{\em Appendix}---.  Here we prove the two lemmas that were used earlier.

\noindent{\bf Lemma 1}
\emph{(Concavity of the Helstrom limit)
Consider the problem of discriminating between states $\hat{\sigma}_0=\int\!{\rm d} {\bm x}\, f_{\bm X}({\bm x}) \hat{\rho}_0({\bm x})$ and $\hat{\sigma}_1=\int\! {\rm d} {\bm x}\, f_{\bm X}({\bm x}) \hat{\rho}_1({\bm x})$, where ${\bm X}$ is a random vector, that occur with prior probabilities $\pi_0$ and $\pi_1$. The Helstrom limit for this binary state-discrimination task satisfies
$
D_{\pi_0}(\hat{\sigma}_0,\hat{\sigma}_1)
\ge \int\!{\rm d} {\bm x}\,  f_{\bm X}({\bm x}) D_{\pi_0}(\hat{\rho}_0({\bm x}),\hat{\rho}_1({\bm x})).$}

\noindent{\bf Proof.}
Let $\hat{M}_0$ and $\hat{M}_1 = \hat{I}-\hat{M}_0$ be the Helstrom-limit positive operator-valued measurement for discriminating between $\hat{\sigma}_0$ and $\hat{\sigma}_1$ when those states' prior probabilities are $\pi_0$ and $\pi_1$.  Then we have that
\begin{eqnarray}
\lefteqn{D_{\pi_0}(\hat{\sigma}_0,\hat{\sigma}_1) = \pi_0 {\rm tr}(\hat{M}_1\hat{\sigma}_0)+\pi_1{\rm tr}(\hat{M}_0\hat{\sigma}_1)} \nonumber \\ 
&=& \int\!{\rm d} {\bm x}\,  f_{\bm X}({\bm x})
\{
\pi_0 {\rm tr}[\hat{M}_1\hat{\rho}_0({\bm x})] + \pi_1{\rm tr}[\hat{M}_0\hat{\rho}_1({\bm x})]\} \nonumber
\\
&\ge& \int\!{\rm d} {\bm x}\,  f_{\bm X}({\bm x}) D_{\pi_0}(\hat{\rho}_0({\bm x}),\hat{\rho}_1({\bm x})), \nonumber
\end{eqnarray}
and the proof is complete.

\noindent{\bf Lemma 2}
\emph{(Error-probability exponent for QI with Rayleigh fading) For $h=0,1$, let $\hat{\bp}_h(\sqrt{\kappa},\phi) = \otimes_{m=1}^M\hat{\rho}_h^{(m)}(\sqrt{\kappa},\phi)$, where $\hat{\rho}_h^{(m)}(\sqrt{\kappa},\phi)$ is the two-mode, zero-mean, Gaussian state whose Wigner covariance matrix is given by Eq.~(\ref{hk}), and let $\hat{\bar{\bp}}_h$ be the unconditional density operators obtained by averaging $\hat{\bp}_h(\sqrt{\kappa},\phi)$ over  Rayleigh and uniform probability density functions for $\sqrt{\kappa}$ and $\phi$, respectively.
Then, for all $\pi_0\pi_1 \neq 0$ we have $\xi_{\rm QI}\equiv-\lim_{M\to\infty}{\ln[D_{\pi_0}(\hat{\bar{\bp}}_0,\hat{\bar{\bp}}_1)]}/M = 0$.}  

\noindent{\bf Proof.}
Because $\kappa \le 1$ is required for a passive target, i.e., one that only reflects, the Rayleigh pdf is really an approximation to $f_{\sqrt{\kappa}}(x) = 2xe^{-x^2/\bar{\kappa}}/\bar{\kappa}(1-e^{-1/\bar{\kappa}})$ for $0\le x\le 1$ that is very accurate in QI target detection's $\bar{\kappa} \ll 1$ scenario.  For proving Lemma~2, however, we need to employ the truncated pdf, so that Lemma~1 and the QCB's exponential tightness for $M$-copy state discrimination gives us
\begin{eqnarray}
\lefteqn{D_{\pi_0}(\hat{\bar{\bp}}_0,\hat{\bar{\bp}}_1)} \nonumber \\
&\ge& \int_0^1\!{\rm d}x\!\int_0^{2\pi}\!{\rm d}y\, \frac{2xe^{-x^2/\bar{\kappa}}}{2\pi \bar{\kappa}(1-e^{-1/\bar{\kappa}})}D_{\pi_0}(\hat{\bp}_0(x,y),\hat{\bp}_1(x,y)) \nonumber\\
&\ge& \int_0^1\!{\rm d}x\!\int_0^{2\pi}\!{\rm d}y\, \frac{2xe^{-x^2/\bar{\kappa}}}{2\pi \bar{\kappa}(1-e^{-1/\bar{\kappa}})} \nonumber \\
&&\hspace{.2in}\times\,\,C_{x,y}(M) e^{-M\xi_{\rm QCB}(\hat{\rho}_0(x,y),\hat{\rho}_1(x,y))}, \nonumber
\end{eqnarray}
where the subunity prefactor, $C_{x,y}(M)$, is an algebraic function of $M$.  Specifically, for all $0\le x\le 1$ and $0\le y\le 2\pi$, we have  $\lim_{M\to\infty}\ln[C_{x,y}(M)]/M=0$. It follows that for every $\epsilon>0$ there is a finite $M_\epsilon(x,y)$ such that $C_{x,y}(M)\ge e^{-\epsilon M_\epsilon(x,y)}$ for all $M > M_\epsilon(x,y)$.  

Because $\Omega \equiv \{0\le x\le 1, 0\le y\le 2\pi\}$ is a compact region, there is a \emph{finite} $M^\star_\epsilon = \max_{(x,y)\in \Omega} M_\epsilon(x,y)$.  So, for all $M > M^\star_\epsilon$ we have
\begin{eqnarray}
D_{\pi_0}(\hat{\bar{\bp}}_0,\hat{\bar{\bp}}_1) &\ge&  e^{-\epsilon M}\int_0^1\!{\rm d}x\!\int_0^{2\pi}\!{\rm d}y\, \frac{2xe^{-x^2/\bar{\kappa}}}{2\pi \bar{\kappa}(1-e^{-1/\bar{\kappa}})} \nonumber \\
&& \hspace{.2in}\times\,\,e^{-M\xi_{\rm QCB}(\hat{\rho}_0(x,y),\hat{\rho}_1(x,y))} \nonumber
\end{eqnarray}
But $\min_{(x,y)\in \Omega}\xi_{\rm QCB}(\hat{\rho}_0(x,y),\hat{\rho}_1(x,y))$ occurs at $x=0$, where $\xi_{\rm QCB}(\hat{\rho}_0(0,y),\hat{\rho}_1(0,y)) = 0$, because $\hat{\bar{\bp}}_0 = \hat{\bar{\bp}}_1$ when the target return's intensity vanishes.  Thus, for any $0< \epsilon'<1$ we can define $\Omega_{\epsilon'} = \{(\sqrt{\kappa},\phi) : \xi_{\rm QCB}(\hat{\rho}_0(x,y),\hat{\rho}_1(x,y)) \le \epsilon'\}$, and then weaken our previous lower bound on the Helstrom limit to 
\begin{equation}
D_{\pi_0}(\hat{\bar{\bp}}_0,\hat{\bar{\bp}}_1)  \ge e^{-(\epsilon+\epsilon') M}\Pr[(\sqrt{\kappa},\phi)\in \Omega_{\epsilon'}] > 0,\nonumber
\end{equation}
where the last inequality follows from $\pi_0\pi_1 \neq 0$.

Applying this bound to the error-probability exponent then leads to 
\begin{equation}
\xi_{\rm QI}(\hat{\sigma}_0,\hat{\sigma}_1)
\equiv-\lim_{M\to\infty}{\ln[D_{\pi_0}(\hat{\bar{\bp}}_0,\hat{\bar{\bp}}_1)]}/M\nonumber \\
\le \epsilon + \epsilon'
\nonumber
\end{equation}
Because this upper bound holds for all $\epsilon, \epsilon' > 0$, by continuity our proof is now complete.


\begin{thebibliography}{99}
\bibitem{footnote2}These prior analyses are easily adapted to known $\kappa$ and known $\phi$, but in all that follows we will let $\phi = 0$ serve as a proxy for an arbitrary known phase value.
\bibitem{footnote3}An alternative SNR calculation for Rayleigh-fading QI with OPA reception that exploits the ${\rm Var}_1(\hat{N})$ signature of target presence yields an even lower value, ${\rm SNR}_{\rm OPA} \simeq (G-1)(\bar{\kappa} N_S)^2/N_B$, when $M\bar{\kappa}N _S/N_B \gg 1$ and $(G-1)N_B \ll 1$.
\end{thebibliography}
\end{document}